# Recovery of surface reflectance spectra and evaluation of the optical depth of aerosols in the near-IR using a Monte-Carlo approach: Application to the OMEGA observations of high latitude regions of Mars.

Mathieu Vincendon[1], Yves Langevin[1], François Poulet[1], Jean-Pierre Bibring[1], Brigitte Gondet[1]

[1]*Institut d'Astrophysique Spatiale, CNRS/Université Paris Sud, Orsay, France.*

**Abstract:** We present a model of radiative transfer through atmospheric particles based on Monte Carlo methods. This model can be used to analyze and remove the contribution of aerosols in remote sensing observations. We have developed a method to quantify the contribution of atmospheric dust in near-IR spectra of the Martian surface obtained by the OMEGA imaging spectrometer on board Mars Express. Using observations in the nadir pointing mode with significant differences in solar incidence angles, we can infer the optical depth of atmospheric dust, and we can retrieve the surface reflectance spectra free of aerosol contribution. Martian airborne dust properties are discussed and constrained from previous studies and OMEGA data. We have tested our method on a region at 90°E and 77°N extensively covered by OMEGA, where significant variations of the albedo of ice patches in the visible have been reported. The consistency between reflectance spectra of ice-covered and ice-free regions recovered at different incidence angles validates our approach. The optical depth of aerosols varies by a factor 3 in this region during the summer of Martian year 27. The observed brightening of ice patches does not result from frost deposition but from a decrease in the dust contamination of surface ice and (to a lower extent) from a decrease in the optical thickness of atmospheric dust. Our Monte Carlo–based model can be applied to recover the spectral reflectance characteristics of the surface from OMEGA spectral imaging data when the optical thickness of aerosols can be evaluated. It could prove useful for processing image cubes from the Compact Reconnaissance Imaging Spectrometer for Mars (CRISM) on board the Mars Reconnaissance Orbiter (MRO).

## 1 Introduction

The highly variable optical thickness of aerosols in the Martian atmosphere has a major impact on the observed reflectance factor at visible, near IR and thermal IR wavelengths. Remote sensing observations of Mars can therefore be used to study aerosols properties such as size distribution or composition, and the contribution of aerosols has to be taken into account when interpreting measured reflectances in terms of surface properties. Emission phase function sequences have been widely used to retrieve scattering properties of aerosols and albedos of the surface in the broad band channel centered on 0.67 μm of IRTM [Clancy and Lee, 1991] and then TES [Clancy et al., 2003]. These studies use the discrete ordinate method DISORT [Stamnes et al., 1988] to model radiative transfer in the atmosphere. Another widely used model of radiative transfer is the SHDOM code by Evans [1998]. However, the SHDOM site does mention that Monte Carlo methods are faster and more accurate for simulations in which there are relatively few modeled radiative quantities. Erard et al. [1994] have studied the effect of aerosols on the observation of the surface in the near-IR. The model of aerosols presented in their study relies on a single scattering approximation, and absorption by aerosols is not considered. This model is therefore mainly valid for low integrated optical thicknesses from the Sun to the surface, then the instrument. Paige et al. [1994] have modeled the effect of aerosols on IRTM measurement above the north polar cap, but the effectiveness of the model was hampered by uncertainties in the selected optical properties of Martian aerosols.

The visible and near-IR imaging spectrometer OMEGA, on board the Mars Express Mission, has provided an extensive coverage of the northern high latitudes of Mars during the summer of 2004. The spectral range of OMEGA (0.35 µm–5.1 µm), in particular the near-IR range (1 µm– 2.5 µm) provides unambiguous information on the presence of water ice at the surface from the strong absorption features at 1.25, 1.5 and 2 µm [Bibring et al., 2004a, 2004b]. Furthermore, the relative strength of these absorption features depends on the path length of photons within water ice. Discriminating the reflectance spectra of icecovered regions from aerosol contributions is therefore important for constraining the texture and the dust contamination of the ice.

We have developed a model of radiative transfer in a layer of atmospheric particles based on a Monte Carlo approach which can be used to analyze the contribution of aerosols in remote sensing observations. We present a method for quantifying and removing the contribution of aerosols to the surface reflectance measured by OMEGA on-board Mars Express, using this model. This method relies on observations at different incidence angles for evaluating the optical thickness of aerosols. The choice of the Monte Carlo approach is justified since we consider only a few input and output parameters. The Monte Carlo model is also well suited for taking into account multiple scattering, which plays a major role at high incidence angle or when the optical thickness of aerosols is high. We take into account the improved understanding of the optical parameters of Martian aerosols provided by recent studies [Ockert-Bell et al., 1997; Clancy et al., 2003; Johnson et al., 2003; Tomasko et al., 1999] and observations by OMEGA. Most OMEGA observations use the nadir pointing mode; hence at this stage the model is dedicated to this observation geometry.

This method has been tested on a northern ice-filled crater of Mars (77°N, 90°E) during the summer of Mars Year 27 (we use the Mars years numbering of Clancy et al. [2000] beginning 11 April 1955). Significant variations of the visible albedo have been observed in this region during previous summers [Bass et al., 2000; Hale et al., 2005], as well as variations of the water ice bands depth in the near-IR range during Mars year 27 [Langevin et al., 2005a]. These changes have been attributed to an evolution of surface properties with diverging conclusions: Bass et al. [2000] explained the brightening of the crater during summer by the deposition of water frost whereas Langevin et al. [2005a] modeled the evolution of the near-IR spectrum during summer by the sublimation of surface water frost and by a removal of surface dust. Significant variations of the optical thickness of atmospheric dust at 9 µm have been reported at 77 °N during the summer of Martian year 25 and 26 [Smith, 2004]. Numerous local and regional dust storms have been observed by MOC in these latitudes during the 1999 north summer [Cantor et al., 2001]. It is therefore of interest to study the contribution of aerosols in the apparent reflectance evolution of this region so as to analyze their impact on inferred surface characteristics. Such a study can be performed using OMEGA as this region has been extensively observed at close time intervals with large variations in solar incidence angles.

In section 2 we present the radiative transfer model which is based on a Monte Carlo approach and the optical properties of aerosols and surface material selected as input parameters. In section 3 we present our method for separating surface and atmospheric dust contribution. We then apply it to the OMEGA observations of a region containing both icy and ice-free surfaces during northern summer.

## 2 A Monte Carlo model for assessing the impact of aerosols on the observed spectra

### 2.1 Model Simulations

For each wavelength, the model simulates the path of a large number of test photons. We consider a layer of identical atmospheric particles above a surface with no geographic variations in surface or aerosol properties. The inputs corresponding to the aerosols layer are: the optical depth t at normal incidence, the phase function of aerosols and the single scattering albedo w. The others inputs are the photometric function of the surface, the albedo A of the surface and the solar incidence angle i. The fate of each test photon is determined as follows:

1) We first draw a random number x between 0 and 1. If this number is higher than $\exp(-\tau/\cos(i))$, the test photon interacts with aerosols before reaching the surface. The level at which the interaction occurs corresponds to a remaining optical depth $\tau' = -\cos(i)\log(x)$.
2) If the photon interacts with an atmospheric particle, we draw a second random number and we compare it to ω to determine whether the photon is absorbed or scattered.
3) If the photon is scattered, we determine its new propagation direction by a random selection compliant with the probability distribution of the phase function.
4) The process is repeated from step 1 taking into account the new propagation direction and the remaining optical thickness between the photon and the surface (if scattered downward) or between the photon and free space (if scattered upward). Each test photon eventually escapes to free space, reaches the surface or is absorbed by aerosols.
5) If the test photon reaches the surface we compare a random number with A so as to determine whether it is absorbed or scattered by the surface.
6) If the photon is scattered by the surface, we determine in which direction by a random selection compliant with the scattering law of the surface. The photon then moves up into the aerosol layer, and we go back to step 1 with the new direction.
7) Eventually, each test photon escapes to free space (in which case the radiance is incremented in the corresponding direction), is absorbed by the surface or by aerosols.

We run this simulation for ~ 10 million photons and we determine the reflectance factor (RF, which corresponds to I/F divided by the cosine of the solar incidence) as observed in a nadir pointing mode as follows: we compute the number of photons as a function of the emergence angle by summing all azimuth angles. We then fit a Lambert law to the observed distribution for small emergence angles (i.e., as seen in nadir pointing mode). An example of this procedure is provided in Figure 1. We have compared our Monte Carlo model with a simple algorithm by F. Forget et al. [2007] which provides a high computational speed result very close to that of the SHDOM radiative transfer code [Evans, 1998] but only for small optical thicknesses. Our results are very similar in the low-opacity domain (Figure 2) which validates the Monte Carlo model.

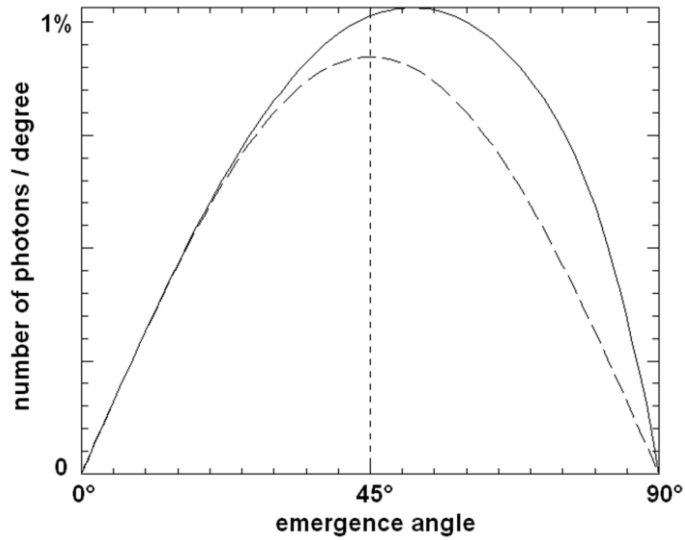

*Figure 1. Distribution of photons which escape the layer of aerosols as a function of emergence (e = 0°: zenith). The surface is supposed to be Lambertian (albedo $A_L$ = 0.6: icy surface), and the phase function of aerosols is a single-lobed Henyey-Greenstein function (asymmetry parameter g = 0.63). The other inputs are t = 0.4, w = 0.97, and i = 70°. The solid line represents the result of the model. It is fitted with a Lambert law (~cos(e)sin(e)) for angles for determining the reflectance factor observed by an instrument in nadir pointing mode (dashed line: RF = 0.53). It is lower than that expected from the proportion of photons escaping into free space (63%), which is itself slightly higher than that which would be observed if there was no aerosol layer (60%). In this example, 5% of photons are absorbed by aerosols, 32% are absorbed by the surface, 42% undergo multiple scattering by aerosols, and 3% undergo multiple scattering by the surface.*

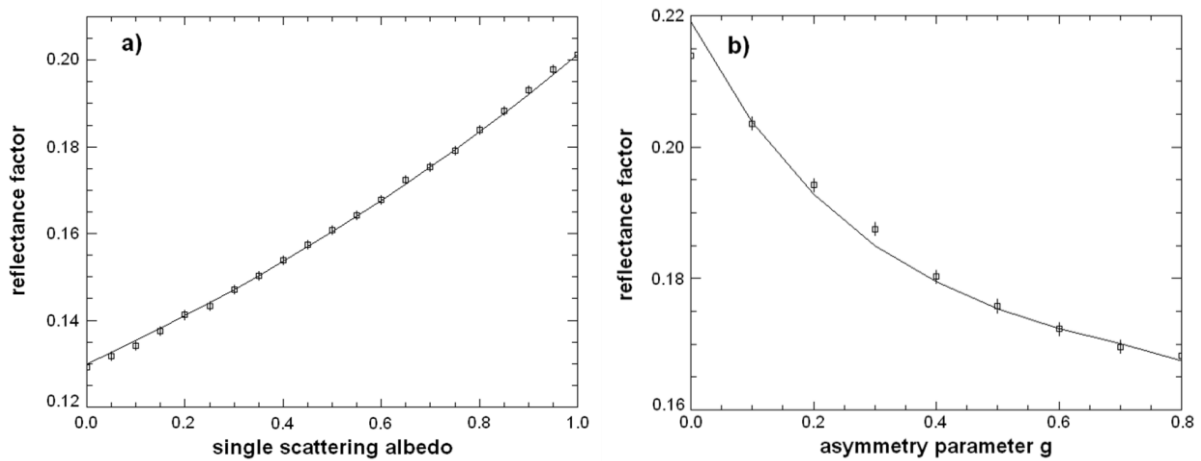

*Figure 2. Comparison between the Monte Carlo model (squares) and a model by Forget et al. (2007) (solid lines) valid at low optical depths. The surface is supposed to be Lambertian (albedo $A_L$ = 0.2), and the phase function of aerosols is a single-lobed Henyey- Greenstein function (asymmetry parameter g). The solar incidence angle is 30 . The reflectance factor in the nadir pointing mode is computed for (a) different values of the single scattering albedo w (t = 0.2 and g = 0.63) and (b) different values of g (t = 0.4 and w = 0.8). Error bars of the Monte Carlo model for $2.10^7$ test photons are indicated. The two models give similar results.*

## 2.2 Selected properties of dust aerosols: phase function

The Monte Carlo model can be applied to any phase function. A detailed study of the optical properties of atmospheric dust in the near-IR is that of Ockert-Bell et al. [1997]. Assuming a Henyey-Greenstein phase function, their best estimate of the asymmetry parameter is a constant value of g = 0.63 between 1 and 2.5 µm. This value of g is consistent with the 0.64 to 0.67 range derived by Clancy et al. [2003] from an analysis of the [0.3 µm–3 µm] broad band channel of TES, modeled as a single wavelength channel at approximately 0.7 µm. As the OMEGA observations of interest for ices are obtained in the 1 to 2.7 µm range, we have selected the phase function derived by Ockert-Bell et al. [1997]. Many studies of the phase function have been achieved for wavelengths shorter than 1 µm [e.g., Tomasko et al., 1999; Markiewicz et al., 1999]. We analyze in section 3.3 the influence of the selected phase function on our method by considering the phase function derived by Tomasko et al. [1999] for a wavelength of 0.965 µm.

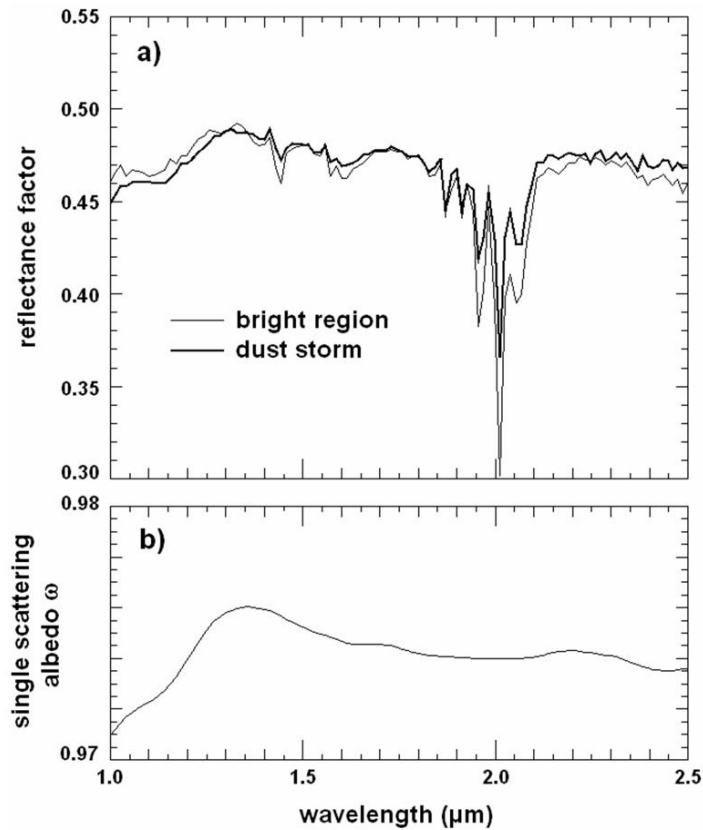

*Figure 3. Determination of the single scattering albedo of atmospheric dust (ω) as a function of wavelength. (a) Thinsolid line: reflectance spectrum of a bright region (256.6°E, 7.6°N, $L_S$ 170.8°). The MOLA altitude is 4.9 km. Thick solid line: reflectance spectrum of a dust storm (24.6°E, 2.4°S, $L_S$ 135.6°). The MOLA altitude is 1.5 km. Spectra are not corrected for atmospheric absorptions. The weak atmospheric $CO_2$ absorption bands for the $L_S$ = 135.6° observation indicate that most photons are scattered high in the atmosphere. This spectrum is therefore representative of a layer of atmospheric dust with a large optical thickness. It is similar to that of the brightest regions of Mars. (b) Values of w inferred from this observation with the Monte Carlo model assuming a Henyey-Greenstein phase function with g = 0.63 (uncertainty ±0.5%).*

### 2.3 Selected properties of dust aerosols: single scattering albedo

For a given particle phase function, the reflectance of an aerosols layer with an infinite optical depth depends only on the single scattering albedo ω of aerosols, which defines the fraction of photons which are absorbed for each interaction with aerosols. In their study of dust parameters, Ockert-Bell et al. [1997] assumed that atmospheric dust is spectrally similar to bright regions. They have used as a reference a spectrum of Amazonia Planitia obtained by ISM. The mean reflectance factor is 0.35 between 1 and 2.5 μm, from which they concluded that ω varies slowly around 0.955 in this wavelength range. With our Monte Carlo model, we indeed obtain a reflectance factor of 0.35 for a layer of infinite optical depth with ω = 0.955 and g = 0.63. Ockert-Bell et al. [1997] mentioned that well calibrated dust storm spectra which were then not available would be more representative of airborne dust. OMEGA observed such an optically thick local dust storm. The reflectance factor is 0.45 at 1 μm (Figure 3a). This value is consistent with the highest albedos observed by OMEGA in the near-IR at scales of a few hundred meters. The value of ω(λ) which can be derived from the observation of the dust storm ranges from 0.971 to 0.976 in the 1–2.5 μm wavelength range (Figure 3b), with an uncertainty of ±0.5% due the uncertainty on the reflectance level of the dust storm spectrum. These values are at the edge of the 2% error of Ockert-Bell et al. [1997]. The value of w at 1 μm (0.971) is consistent with that estimated by Johnson et al. [2003] at 1 μm: 0.969.

### 2.4 Selected photometric function for the surface

Our study focuses on nadir pointing observations with large incidence angles (hence potentially large aerosol contributions). The phase angle varies between 55° and 77° (Table 1). Johnson et al. [2006] have observed and modeled the photometric function of the surface at 1 μm for different rocks and soils observed by Pancam/Spirit. For the intermediate phase angles of interest, the best fit with the observed bidirectional reflectance distribution function (BDRF) using the theory of Hapke [1993] results in relative variations of at most 10% when compared to a constant BDRF (Lambert scattering). In contrast, aerosols scattering can increase the reflectance factor of dark surfaces by up to 50% in relative terms (Figure 11). We can therefore approximate the actual surface photometric function by a Lambert hypothesis as we consider moderate phase angles and dark surfaces. This hypothesis will be tested in section 3.2 when comparing recovered surface spectra at different incidence angles.

## 3 Application to bright ice deposits within a north polar crater during summer.

### 3.1 Observations

A test case for the impact of aerosols on observed versus actual surface optical properties is provided by a bright, ice-filled crater located at 77°N and 90°E and an intermediate albedo region centered ~50 km to the east of the crater. These regions have been previously observed in the visible, with significant and complex changes in the albedo during summer. Mariner 9 and Viking images of different Martian years in summer have been compared by Bass et al. [2000]. A seasonal brightening of specific regions in the crater area is observed each year during summer until $L_S$ 136, notably between $L_S$ ~ 99° and $L_S$ ~ 117° (see Figure 4). The bright crater bowl (region A) and the region with a middle albedo at the east of the crater (region B) become brighter during early summer. This albedo increase has been attributed to an increase of the extent of $H_2O$ ice coverage due to frost deposition [Bass et al., 2000]. Three Martian years of visible MOC images of the crater have also been studied by Hale et al. [2005]. A possible brightening between $L_S \sim 108°$ and $L_S \sim 140°$ and a darkening between $L_S \sim 140°$ and $L_S \sim 170°$ are observed. MOC images lead the team to conclude that the crater remains consistently frost covered, and that the late summer darkening may be due to a deposition of dust on the ice inside the crater. These diverging conclusions as well as the study of Kieffer [1990] demonstrate that assessing frost deposition and ice evolution from visible data alone is not straightforward.

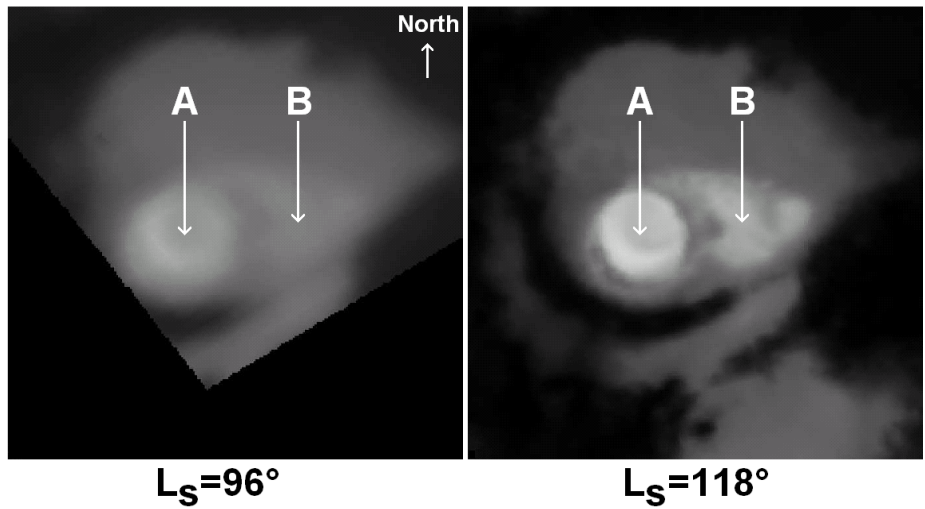

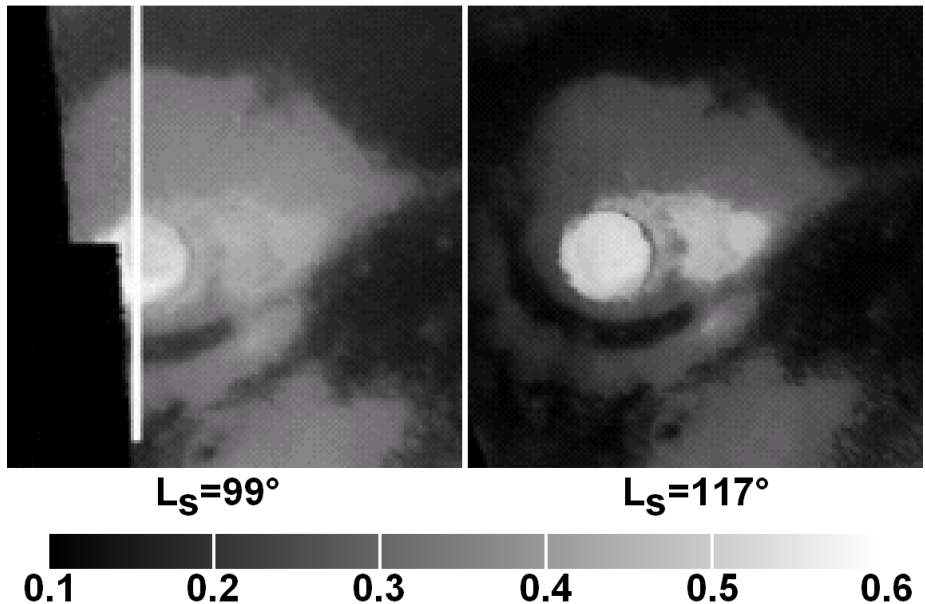

*Figure 4. Images of a crater filled with water ice at 77°N and 90°E. The crater bowl is approximately 30 km in diameter. (a) Reflectance factor observed by OMEGA in 2004 at 1.085 mm (continuum) scaled from 0.1 (black) to 0.6 (white). (b) Reflectance observed by Viking in 1978 in the red filter [Bass et al., 2000]. The albedo evolution during early summer observed by OMEGA in 2004 is similar to that observed by Viking in 1978. Region A (77.09°N, 89.11°E, within the crater) and region B (77.15°N, 91.52°E, in the outlying deposit) both become brighter at $L_S$ 118°, while dark regions become darker.*

*Table 1. List of OMEGA observations of the crater used in this study.*

|   | Cube Number | $L_S$ | Incidence | Emergence | Phase | Altitude, km | Local Time | Observation Date |
|---|---|---|---|---|---|---|---|---|
| 1 | 0907_1 | 95.9° | 54-57° | 5-8° | 59-63° | 3200 | 14H | 03 Oct. 2004 |
| 2 | 0940_2 | 100.1° | 54-57° | 3-8° | 58-62° | 2700 | 14H | 12 Oct. 2004 |
| 3 | 0962_2 | 102.8° | 54-57° | 3-7° | 58-62° | 2500 | 14H | 19 Oct. 2004 |
| 4 | 0973_2 | 104.2° | 54-57° | 2-7° | 57-62° | 2300 | 14H | 22 Oct. 2004 |
| 5 | 1001_1 | 107.8° | 68-71° | 0-5° | 70-73° | 2600 | 5H | 30 Oct. 2004 |
| 6 | 1012_1 | 109.2° | 68-71° | 0-5° | 71-74° | 3500 | 5H | 02 Nov. 2004 |
| 7 | 1017_3 | 109.8° | 54-57° | 0-6° | 57-61° | 1900 | 14H | 03 Nov. 2004 |
| 8 | 1023_1 | 110.6° | 69-72° | 1-6° | 72-75° | 3300 | 5H | 05 Nov. 2004 |
| 9 | 1034_1 | 112.0° | 69-72° | 1-6° | 72-75° | 3200 | 5H | 08 Nov. 2004 |
| 10 | 1050_3 | 114.0° | 55-58° | 0-5° | 56-60° | 1600 | 14H | 12 Nov. 2004 |
| 11 | 1056_2 | 114.8° | 70-73° | 2-7° | 74-77° | 2900 | 5H | 14 Nov. 2004 |
| 12 | 1072_3 | 116.9° | 55-58° | 0-5° | 55-60° | 1500 | 14H | 18 Nov. 2004 |
| 13 | 1083_3 | 118.3° | 55-58° | 0-5° | 55-60° | 1400 | 14H | 21 Nov. 2004 |
| 14 | 1085_2 | 118.6° | 77-80° | 2-11° | 71-74° | 2600 | 3H | 22 Nov. 2004 |
| 15 | 1420_0 | 165.4° | 71-73° | 0-3° | 72-74° | 340 | 10H | 24 Feb. 2005 |

The crater was observed by OMEGA on fifteen nadir pointing tracks in summer: 14 low-resolution tracks (128 pixels wide, IFOV ~2 to 4 km) from $L_S = 96°$ to $L_S = 119°$ and one high-resolution track (16 pixels wide, IFOV ~ 300 m) at $L_S$ 165°. With the nearly polar orbit of Mars Express, a region at a latitude of 77° could be observed with an incidence angle of ~57° (close to local noon) and ~75° (close to local midnight) in early summer, as this area is then continuously illuminated. The list of observations used in this study is indicated in Table 1. Spectra of the reflectance factor (RF) are corrected for absorptions by atmospheric gases (mainly $CO_2$) with the method used by Langevin et al. [2005a] and Mustard et al. [2005]. Additional information on atmospheric correction of OMEGA spectra are provided by Melchiorri et al. [2006] on the basis of a comprehensive line by line analysis which does not take into account aerosol scattering.

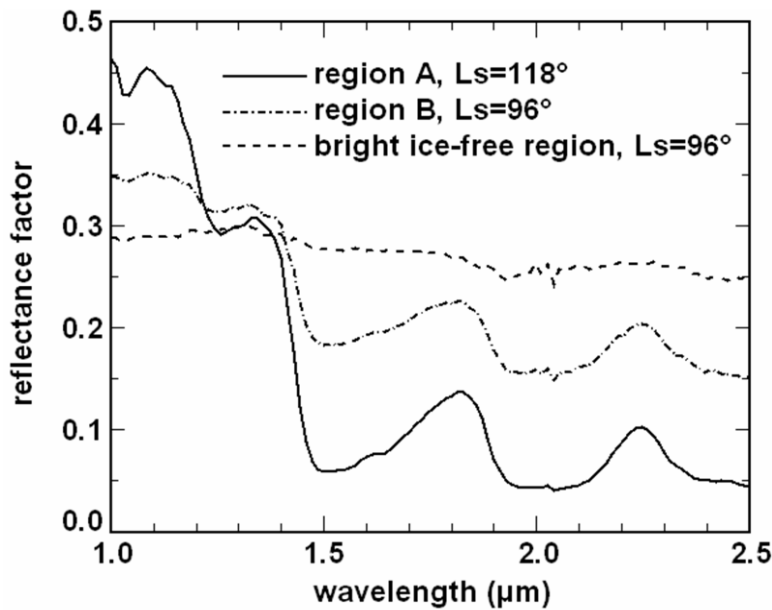

*Figure 5. Near-IR spectra seen by OMEGA in the crater area, which demonstrate that water ice is present in both region A and region B.*

The reflectance factor at 1.085 µm corresponds to a spectral region which is outside the major bands of H$_2$O ice; hence it corresponds to the albedo in the continuum (Figure 5). A significant albedo increase within the crater (region A: from RF = 0.39 to RF = 0.45) and close to the crater (region B: from RF = 0.35 to RF = 0.44) is observed between L$_S$ = 96° and L$_S$ = 118°, in agreement with the result of Bass et al. [2000] during a comparable period (L$_S$ = 97°–117°) of other Martian years. Region B, which has approximately the same albedo as the surrounding terrains at L$_S$ 96°, became brighter and more contrasted at L$_S$ 118°. Typical examples of OMEGA spectra observed in the crater area are shown in Figure 5. Although region B is similar in reflectance at 1 mm to the surrounding regions at L$_S$ 96° (RF ~ 0.35), strong spectral signatures of water ice are observed. The ice boundaries are strikingly stable during the analyzed period (Figure 6). This result shows that the albedo increase of region B in Figure 4 is not due to a deposition of ice frost on an ice-free surface, as proposed by Bass et al. [2000]. The albedo increase in regions A and B is correlated with an albedo decrease in dark ice-free regions (from RF 0.15 to RF 0.10 at 1 mm). Such a behavior is expected if there is a decrease in the optical thickness of dust aerosols [Clancy and Lee, 1991]. Water ice aerosols can be easily detected by OMEGA due to the water ice absorption bands at 1.5 and 2 µm [Gondet et al., 2006]. Transient water ice signatures are observed only in one of the 15 available observations of the region (#12 of table I, at L$_S$ 116.9°, Figure 7). The weak observed spectral features correspond to a small grain size (<5 µm), which is typical of atmospheric water ice [Langevin et al., 2005b, 2007]. If water ice clouds were present above an ice-covered region, discriminating their signatures from that of the surface would not be straightforward from OMEGA data. However, few water ice clouds are observed in early northern summer [Wang and Ingersoll, 2002; Smith, 2004]. We therefore consider in the following section that aerosols are free of water ice.

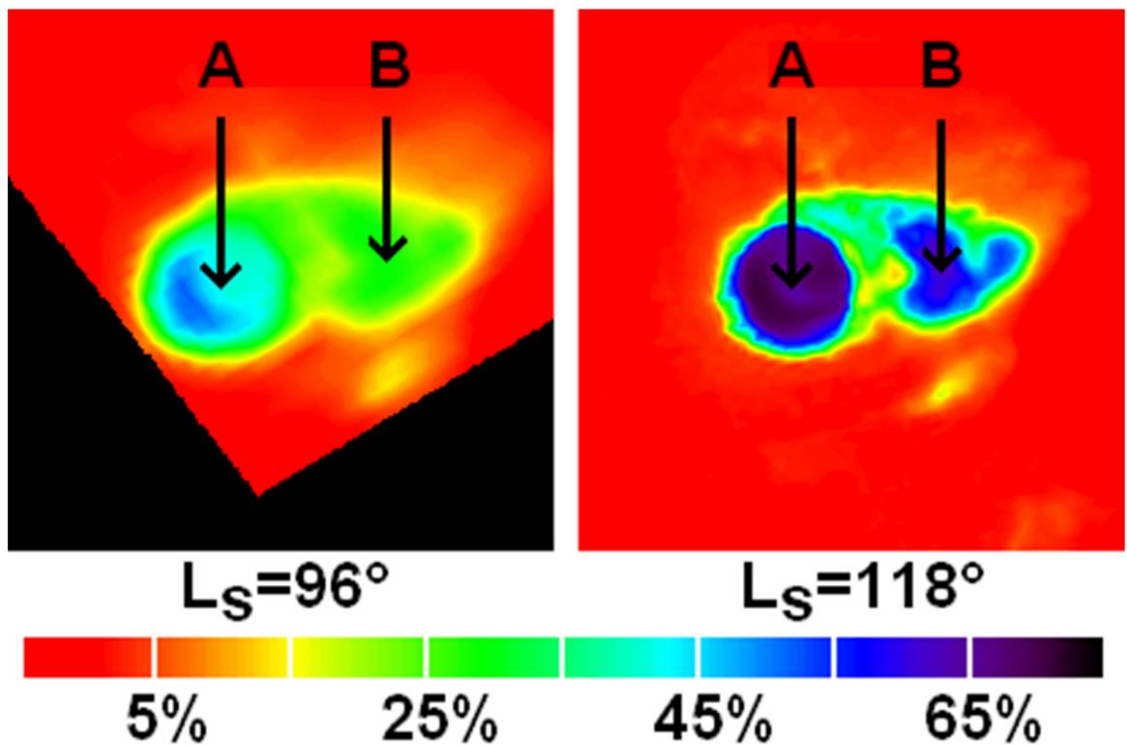

*Figure 6. Map of the strength of the 1.5 µm absorption feature of water ice (red: no ice detected). Images are resampled using a bilinear method. The OMEGA IFOV is two times larger at L$_S$ = 96° (3 km) than at L$_S$ 118° (1.5 km). No increase in the extent of water ice surfaces is observed.*

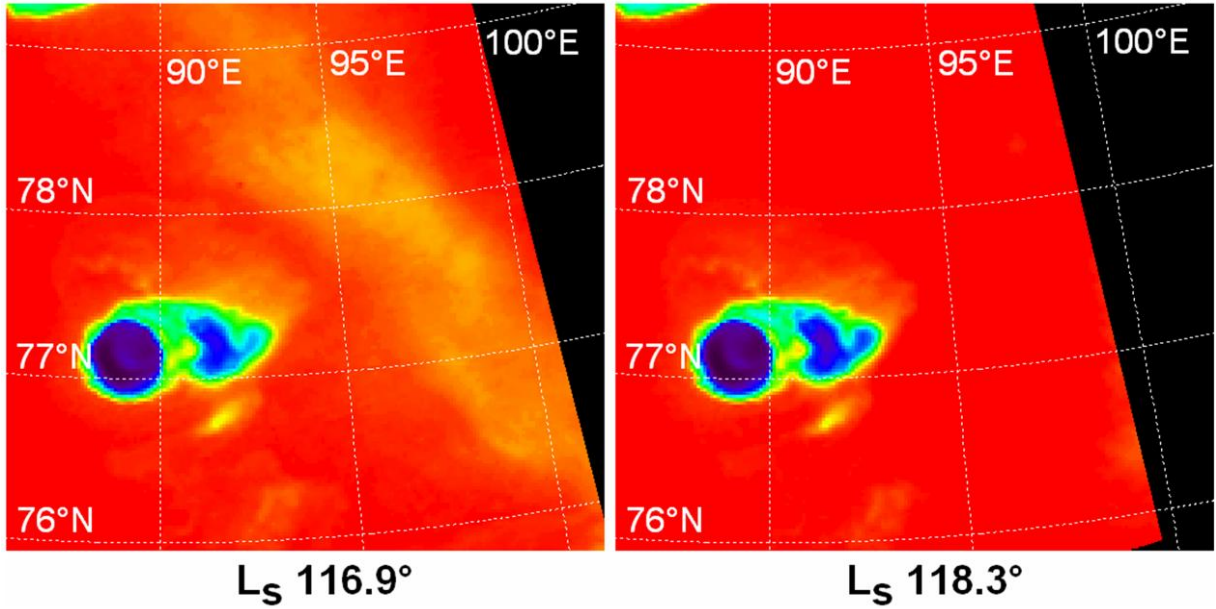

*Figure 7. Weak and transient signatures of $H_2O$ ice appear at $L_S = 116.9°$ and disappear at $L_S = 118.3°$. They are interpreted as resulting from water ice in clouds. The color code is the same as in Figure 6.*

### 3.2 Determination of the spectral characteristics of aerosols and surface

3.2.1 Observations at different incidence angles

From table I, there are 3 sets of observations at different incidence angles obtained within a few days: close to $L_S$ 110° (#6: i=69°; #7: i=57°; #8: i=69°), close to $L_S$ 114° (#10: i=56°; #11: i=71°) and close to $L_S$ 118° ( #13: i=56°; #14: i=78°). For these three values of $L_S$, it is possible to estimate the optical thickness of aerosols at a given wavelength $\tau(\lambda)$ from the evolution the observed reflectance factor of dark regions close to the icy crater, using the results of our Monte-Carlo model (we assume that there are no changes in the surface albedo and in the aerosols optical depth during the short time intervals considered here):

- For each observation, the incidence angle is known (e.g., i=56° for #10). Using the model with the optical parameters as determined in section 2, the observed reflectance factor $RF(\lambda)$ at a given wavelength depends then only on the actual Lambert albedo $A_L(\lambda)$ of the surface and the optical thickness $\tau(\lambda)$ of aerosols.

- The function $RF(A_L, \tau)$ increases monotonically with $A_L$. RF increases monotonically with $\tau$ for low values of $A_L$ ($< 0.25$) and it decreases monotonically with $\tau$ for high values of $A_L$ ($> 0.45$). Except for terrains with intermediate albedos, it is therefore possible to use this relationship so as to obtain $A_L(\tau, RF)$.

- At a given wavelength, the observations at two incidence angles provide two RF, therefore two functions $A_{L1}(\tau)$ and $A_{L2}(\tau)$. The intersection between these two monotonic functions (Figure 8) provides the determination of both $\tau(\lambda)$ and $A_L(\lambda)$.

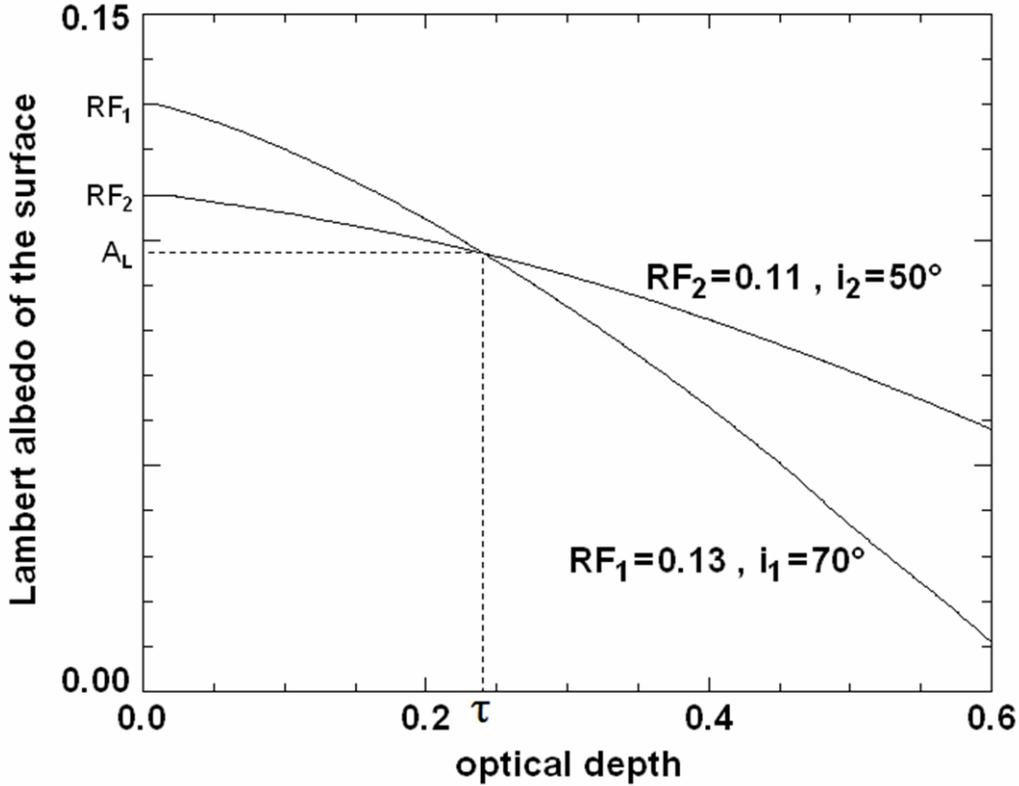

*Figure 8. Determination of the surface Lambertian albedo $A_L$ and the normal optical depth t of the aerosol layer with two observations of the same surface element at different incidence angles, performed a few days apart. There is only one combination of surface Lambertian albedo $A_L$ and optical depth t of the aerosol layer which accounts for the two observed reflectance factors $RF1(i_1)$ and $RF2(i_2)$.*

At $L_S \sim 110°$, the observed reflectance factor spectra for the two solar incidence angles are shown in Figure 9. The two spectra at i=69° are nearly identical, and they are obtained respectively 1.5 days after and before that at i=56°, which validates the assumption that the optical thickness of aerosols has remained nearly constant during this short time period. The resulting values of the surface albedo $A_L(\lambda)$ are also displayed in Figure 9. The variation of the optical thickness $\tau(\lambda)$ with wavelength is shown on Figure 10 and is compared with theoretical models of other authors [Clancy et al., 2003; Drossart et al., 1991]. Our retrieved $\tau(\lambda)$ is compatible with the small mean particle size ($r_{eff}$=1.0 ± 0.2 μm) observed by Clancy et al. [2003] between $L_S$ 50° and $L_S$ 200° in the Northern hemisphere.

Small variations in the observed spectrum $RF(\lambda)$ are amplified in the resulting variation of $\tau$ and $A_L$ with wavelength. These variations correspond to the noise of the instrument and to atmospheric correction artifacts (e.g., near 2 μm). Therefore, we use a linear combination $\tau_0(\lambda)$ of optical depth functions for different particle sizes computed with Mie theory [Clancy et al., 2003, Figure 13] which gives the best fit of the observed wavelength dependence $\tau(\lambda)$ at $L_S \sim 110°$ (Figure 10). The optical depth at a given time can then be determined by searching the value of $\tau(1\ \mu m)$ which minimizes the quadratic spread between surface spectra corrected for aerosol contributions at different incidence angles. For other wavelengths the optical depth is given by $\tau(\lambda) = \tau(1\mu m) \times \tau_0(\lambda)/\tau_0(1\mu m)$.

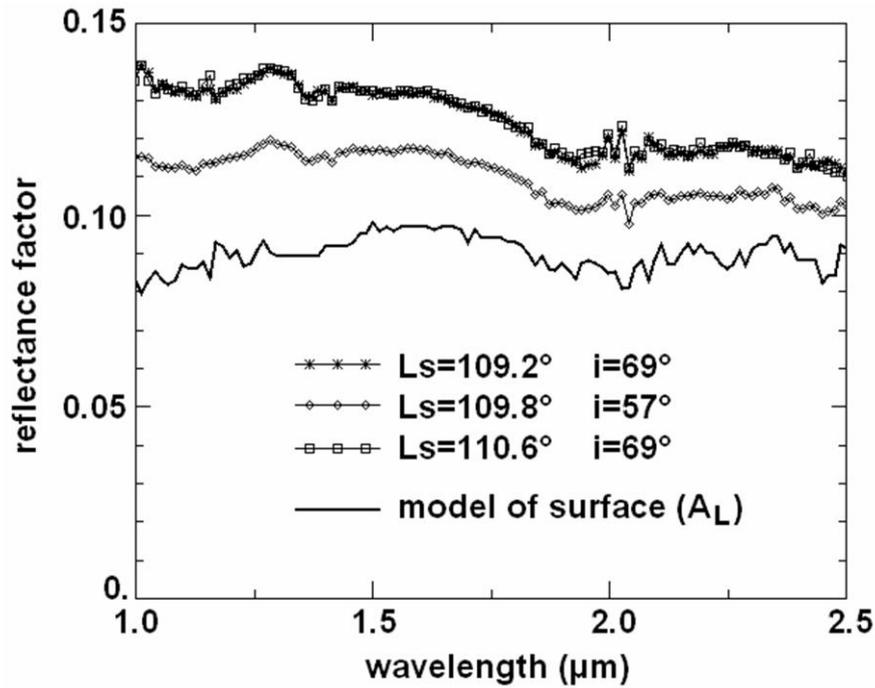

*Figure 9. Consecutive observations of a low-albedo icefree region. The two spectra at the top ($L_S$ 109.2° and $L_S$ 110.6°) have been obtained less than 3 days apart with the same observation geometry. They are very similar, which demonstrates that aerosol properties remained stable during that period. The spectrum obtained at an intermediate time ($L_S$ 109.8°) with a lower solar incidence angle (57° instead of 69°) is markedly different. Using the Monte Carlo model, this can be used for determining the spectrum of the surface (thick solid line) and the optical thickness of aerosols as a function of wavelength (Figure 10).*

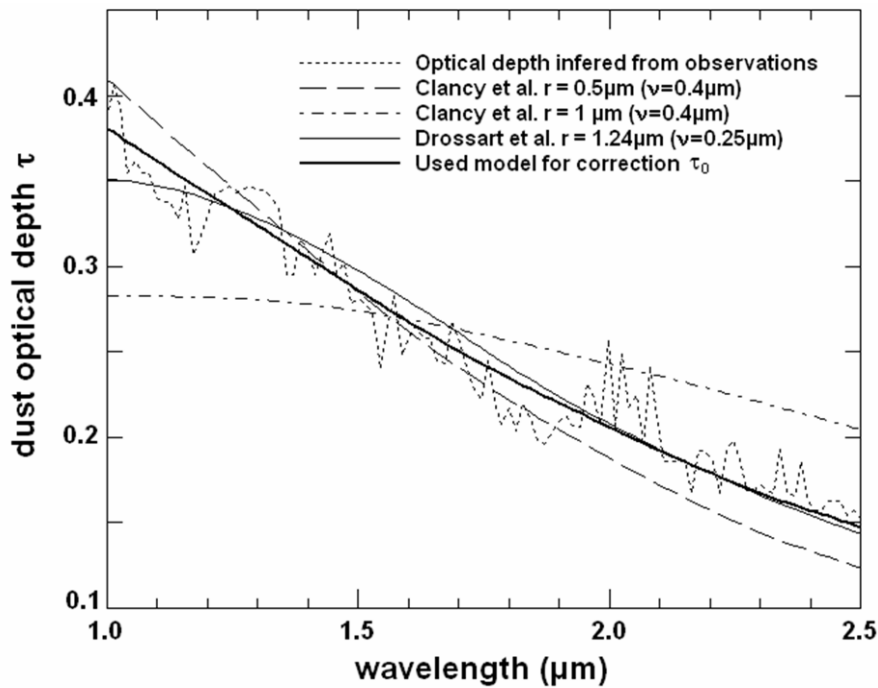

*Figure 10. Optical depth of the aerosol layer as a function of wavelength provided by the model (dotted line) from observations of Figure 9 at $L_S$ = 110° ($\tau(1\ \mu m)$ = 0.38). This*

*variation of τ with λ is compared with Mie calculations of Clancy et al. [2003] for a mean effective radius r = 0.5 μm and r = 1 μm (effective variance veff = 0.4 mm) and with the model of Drossart et al. [1991] for irregular particles with r = 1.25 μm (veff = 0.25 μm). The dependence of optical depth τ 0 with wavelength (thick solid line) used in this study (Figures 11, 12, 13, 14, and 15) corresponds to a linear combination of models of Clancy et al. [2003] which best fits the observation.*

Spectra obtained with i=56° are very similar at $L_S$ 114° (#10), $L_S$ 117° (#12) and $L_S$=118° (#13), therefore the atmospheric dust cover remains stable over this period (Figure 11). We can thus constitute a set of 3 observations with i=56°, 71° and 78° (#10, #11 and #14) for the $L_S$=114°-119° period (Figure 11). With three different incidence angles, the model is strongly constrained. The resulting value of the optical thickness at 1 μm is 0.28 with an estimated uncertainty of ± 0.05 due the noise of the instrument and geographical variations of the optical depth. The satisfactory fit between surface spectra in Figure 11 demonstrates that the wavelength dependence determined at $L_S$ 110° is an adequate model for aerosols at $L_S$ 114°-119°, and that a Lambert scattering behavior for the surface is a good first approximation in our domain of application: moderate phase angles (55°-79°) and dark surfaces where up to 50% of the signal is scattered by aerosols.

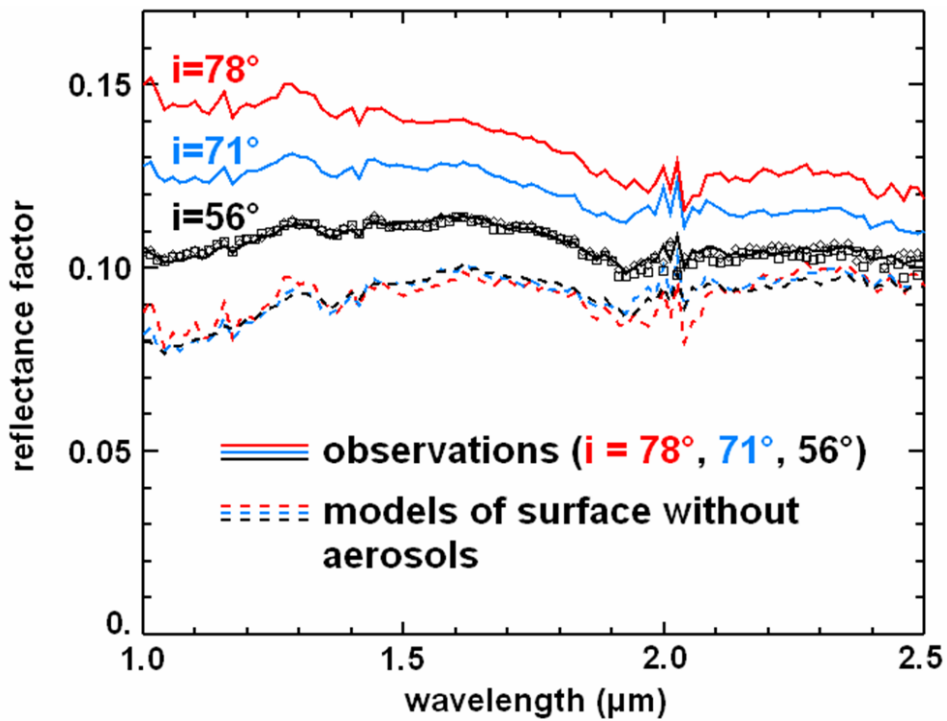

*Figure 11. Three observations of dark areas at $L_S$ 114°– 118° with different incidence angles (solid lines: i = 78°, red; i = 71°, blue; and i = 56°, black). For such a dark surface, the apparent reflectance increases with increasing incidence angle are due to the additive contribution from aerosol scattering. Observations obtained with i = 56° at $L_S$ 114° (#10), $L_S$ 117° (#12), and $L_S$ = 118° (#13) are indicated in black (solid line, square, and diamonds, respectively). The 3 spectra at i = 56° are similar, which indicates a stable atmospheric dust opacity during this period. An optical thickness τ(1 μm) of 0.28 can be determined by minimizing the quadratic spread between the corrected surface reflectance spectra derived from the model. The excellent fit between the three corrected surface reflectance spectra (dotted lines) supports the validity of our aerosol model.*

Aerosols increase the RF of dark regions and decrease the RF of bright regions [Clancy and Lee, 1991]. A major test for our model is the observation with different incidence angles of the spectrum of ice-covered regions, which varies from RF < 0.04 in the absorption bands to RF=0.5 in the continuum. Local slopes have a strong impact on the evolution of the reflectance factor with incidence angle since surface scattering varies with local incidences whereas aerosols scattering varies with global incidences. A flat homogeneous region covered by water ice at the km scale has been selected in the vicinity of the crater (99.8°E, 74.4°S). It has been observed with different incidence angles close to $L_S$ 118° (#13 and #14 in table 1). The inferred optical thickness τ(1 µm) is 0.25, well within our estimated uncertainty of ± 0.05 compared to that inferred above dark areas at the same time (0.28). The resulting surface reflectance spectra (Figure 12) are remarkably similar, which validates our aerosol correction model when strong absorption bands are present.

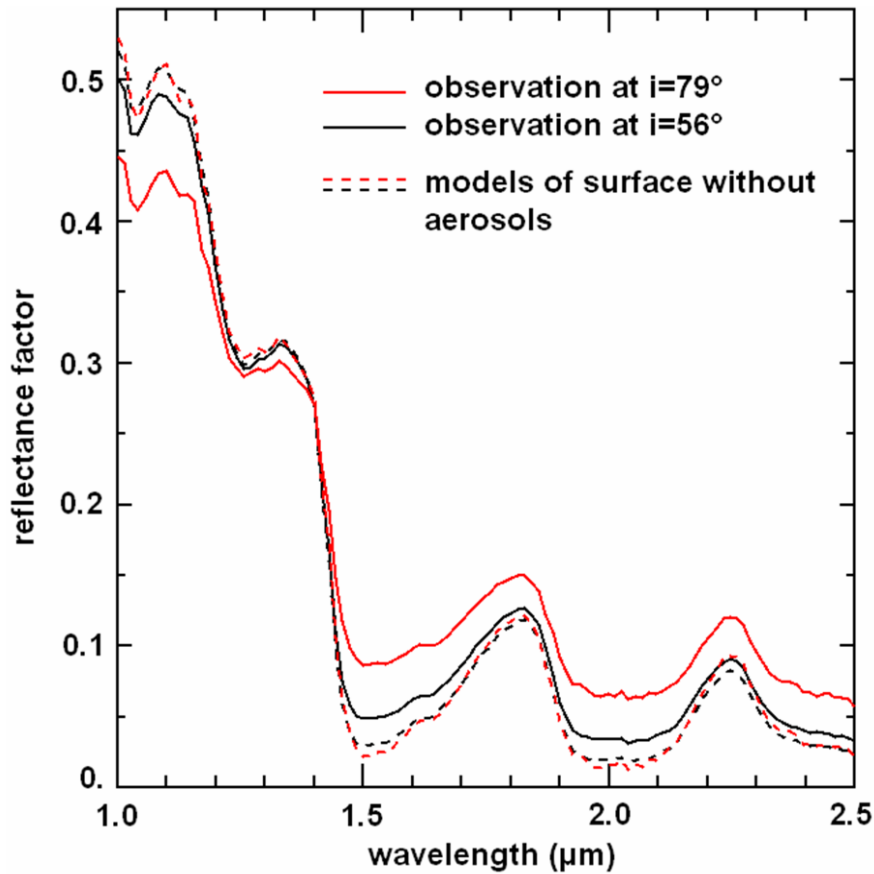

*Figure 12. Similar to Figure 11, but for an ice spectrum. A water ice–covered surface is observed with two different solar incidence angles at $L_S$ 118° (56°, black solid line, and 79°, red solid line). These two spectra give similar surface spectra (dotted lines) after removing aerosol contributions with τ(1 µm) = 0.25.*

### 3.2.2 Evolution with $L_S$ in early summer

At an incidence angle ~ 56°, regions with a low albedo become darker later in the summer ($L_S$ 96° – 119°, Figure 13). This early summer evolution is observed for all dark ice-free regions; hence it can be attributed to a change in the contribution of aerosols. The evolution of the spectrum of dark ice-free regions is presented in Figure 13. At each $L_S$, it is possible to recover a surface spectrum very similar to that inferred at $L_S$ 114-118° by adjusting only one parameter: the optical thickness τ(1µm) of aerosols. The inferred optical thickness at 1µm decreases by a factor of 2.6 from $\tau(1\mu m) \sim 0.73$ at $L_S$ 96° to $\tau(1\mu m) \sim 0.28$ at $L_S$ 114°. Uncertainties mainly result from that on the reference value ($\tau(1\mu m) \sim 0.28$) used to obtained the surface spectrum (typically ±0.05). This range of optical thicknesses is consistent with that measured by Pancam [Lemmon et al., 2004 and 2006] and TES [Clancy et al., 2003] in the Vis / near IR spectral range (0.2 to 1 for most observations). The dust optical depth at 9.3 µm decreases from ~ 0.2 - 0.3 at $L_S$ 90° to ~ 0.05 - 0.1 at $L_S$ 120° during Martian years 25 and 26 for a latitude of 77°N [Smith, 2004]. If we scale these values by a factor 1.25 to take into account the fact that TES measurement are only a measurement of absorption [Clancy et al., 2003; Wolff and Clancy, 2003], we obtain a ratio of ~ 2.5 – 3 between $\tau_{1\mu m}$ (OMEGA) and $\tau_{9\mu m}$ (TES). This ratio is similar to that derived by Clancy et al. [2003] between the visible broadband filter and the 9 µm channel of TES in northern summer. The same variation of the optical depth with wavelength ($\tau_0(\lambda)$ in Figure 10) explained all observations during early summer, which means that no major changes occur in the size distribution or in the composition of aerosols during this period.

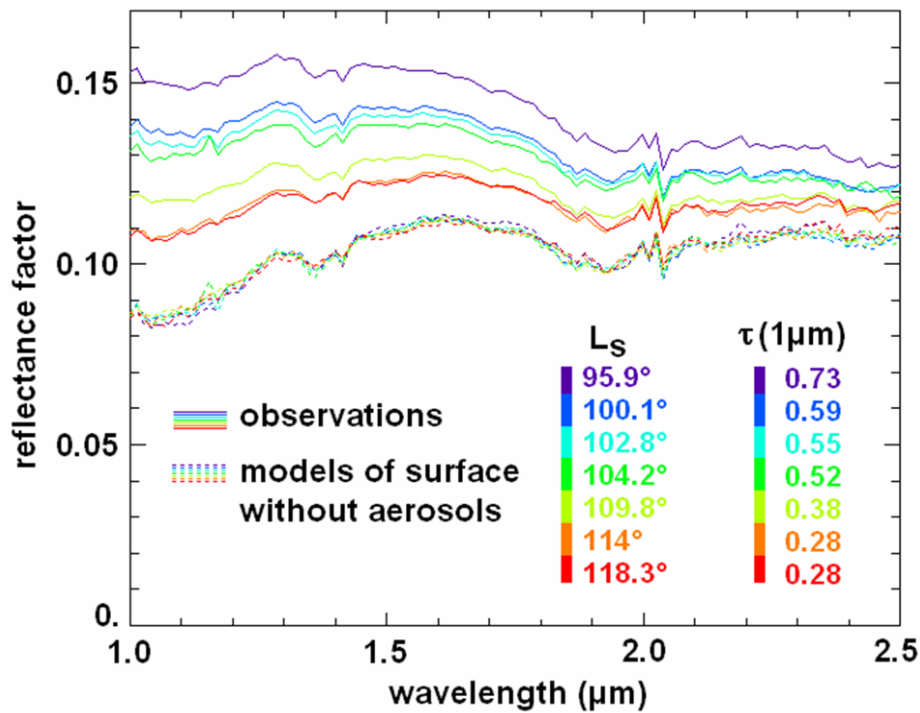

*Figure 13. Evolution of the optical depth as a function of time. Seven observations of dark areas with the same incidence angle (i ~ 56°) show a regular decrease of the reflectance factor for increasing $L_S$ (solid lines). Applying the model provides a consistent surface réflectance spectrum (dotted lines), assuming that the optical depth decreases by a factor 2.6 between $L_S = 96°$ ($\tau$ (1 µm) ~ 0.73) and $L_S = 118°$ ($\tau$(1 mm) ~ 0.28).*

The evolution of the observed reflectance spectrum of the ice covered regions inside the crater in early summer ($L_S$ 96° to $L_S$ 118°) is shown on Figure 14. Assuming that the optical depth inferred above dark surfaces around the crater is similar to that above the crater, we can obtain the surface spectra of the icy region without aerosols contribution. The aerosol optical thickness decreases by a factor 2.6 during this period. However, the evolution of reflectance spectra corrected for aerosol contributions (Figure 14) is qualitatively similar to that of observed spectra. In particular, the increase of the reflectance factor in the continuum (1.08 µm) during this period is not primarily due to the decrease in optical thickness of aerosols. It was attributed by Bass et al. [2000] to the deposition of $H_2O$ ice frost. When $H_2O$ ice seasonal frost is observed by OMEGA [Schmitt et al., 2005], it is fine grained (10 µm to 100 µm) as it is characterized by a weak absorption band at 1.25µm [Grundy and Schmitt, 1998]. The albedo increase is associated with an increase in the 1.25 µm absorption strength, which is not compatible with a deposition of frost.

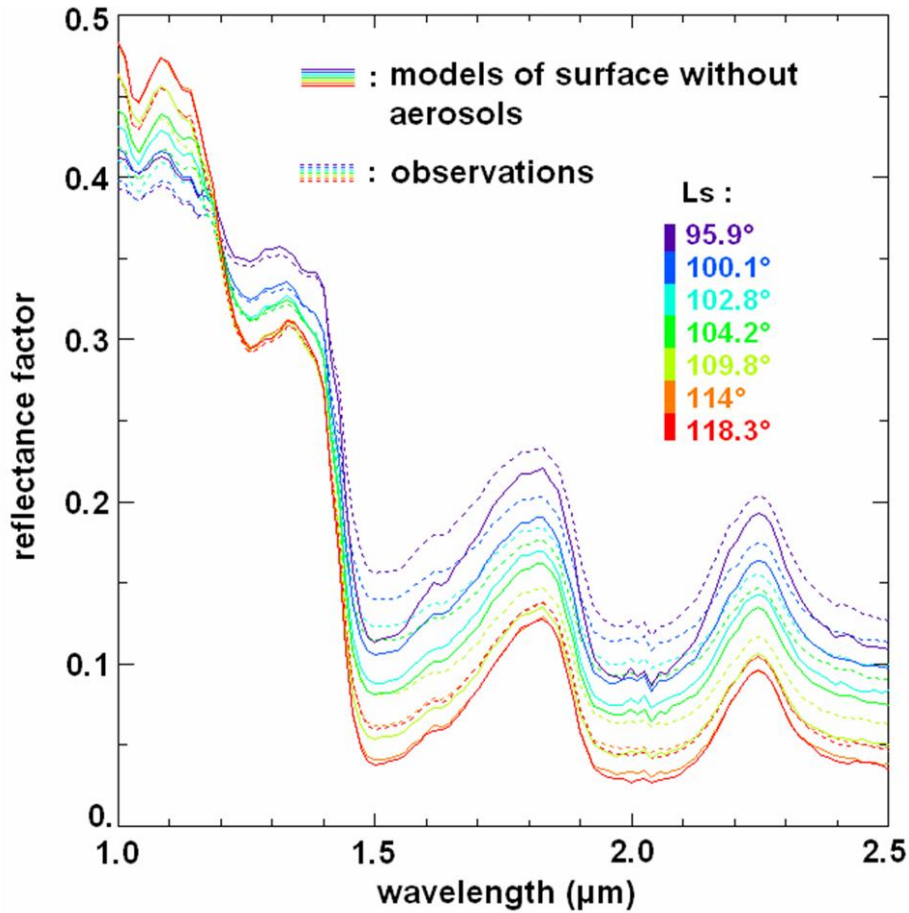

*Figure 14. Evolution of the reflectance spectrum of water ice inside the crater (region A in Figures 4 and 6) after removing aerosol contributions. All spectra are obtained with a similar incidence angle of 56°. Observations correspond to dotted lines, while model-derived surface reflectances are plotted in solid lines. At 1 µm (continuum), only a small fraction of the observed increase of the reflectance factor from $L_S$ 96° to $L_S$ 118° can be attributed to the decrease in the optical thickness of aerosols.*

In order to interpret the observed evolution, we have modelled the corrected spectra with the radiative transfer model of Poulet et al. [2002] which simulated both intimate and intra-mixtures of dust and ice grains of different sizes. The quality of the fits is significantly improved once aerosol correction is applied (Figure 15). Retrieved values for dust contamination and size of ice grains are in part model dependent, but all reasonable fits require a reduction by a factor of two of the dust contamination of surface ice between $L_S$ 96° and $L_S$ 118°. The best fits of the evolution of ice covered regions (Figure 15) are obtained with two populations of ice grains: small grains < 30 µm in size and large grains ~ 900 µm in size, with a decreasing proportion of small grains as a function of time. Our interpretation is that this period corresponds to the last stages of the sublimation of small-grained seasonal water ice frost during early summer, revealing the larger grained permanent ice. The observed decrease in surface contamination by dust grains could be linked to frost sublimation, as proposed by James et al. [1992]. These results obtained after taking into account aerosol effects support the conclusions of Langevin et al. [2005] which attributed the brightening of outer craters to a decrease in dust contamination of predominantly large grained water ice.

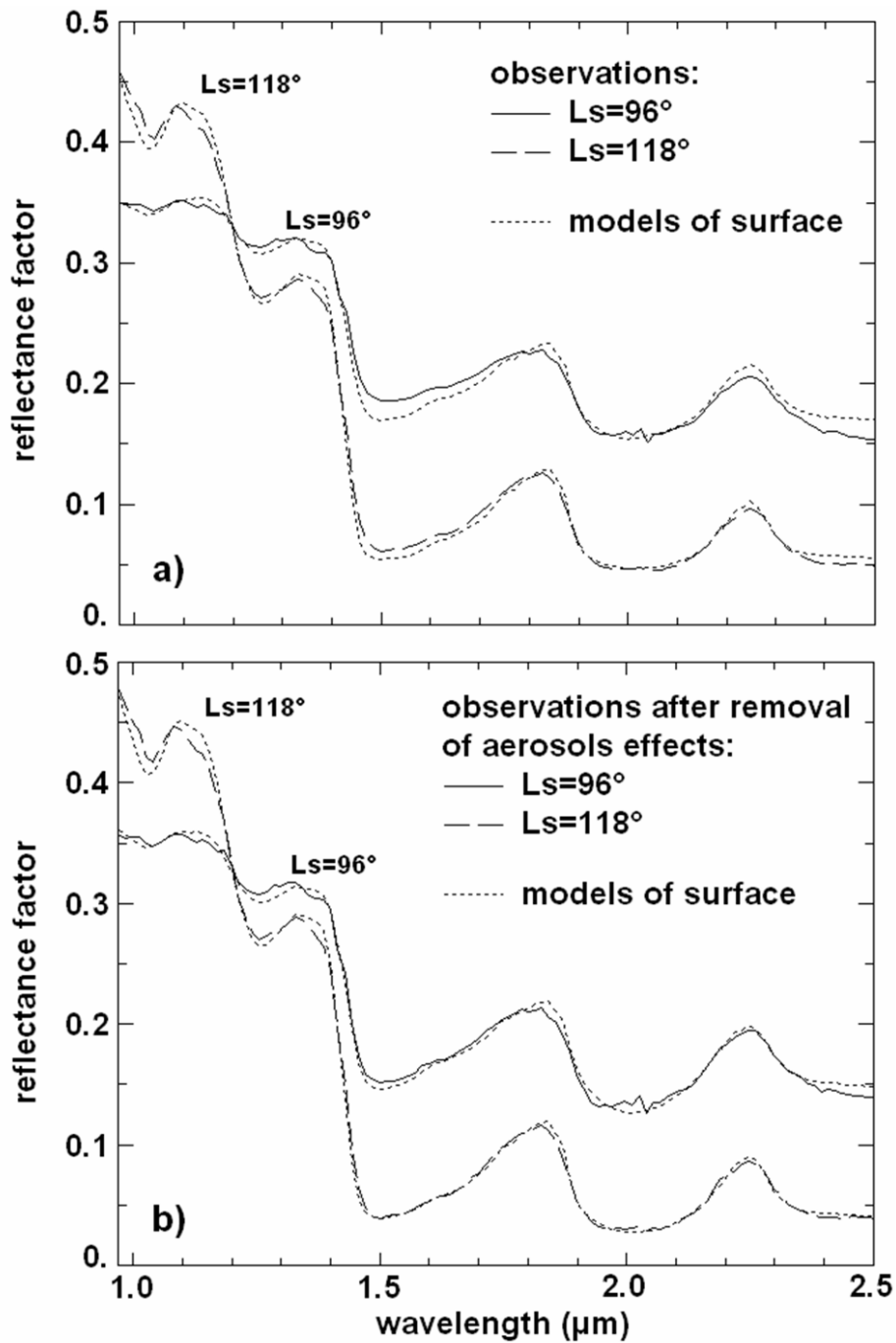

*Figure 15. Examples of models of surface properties for region B of Figure 4. Models assume an intimate mixture of dust and two sizes of ice grains. The model fits of band shapes are significantly improved after the aerosol contribution is removed. The surface is dominated by small water ice grains (30 μm or less) at $L_S = 96°$, whereas large water ice grains (~900 μm) are prominent at $L_S = 118°$. The fraction of dust in volume% within the ice is reduced by a factor of 2 from $L_S$ 96° to $L_S$ 118°.*

### 3.2.3 End of summer: $L_S=165°$

No observations of this region were available from $L_S$ 119° to $L_S$ 165°, at which time a high resolution track was obtained (Figure 16) as the pericenter of Mars Express had moved to high northern latitudes. The observed reflectance spectrum of dark ice-free regions around the crater (Figure 17, bottom) is much brighter at $L_S$ 165° than at $L_S$ 119°. Assuming that actual surface reflectance properties did not change, an optical depth τ(1 μm) of 0.85 is required for aerosols, with a slightly larger aerosols mean particle size than during early summer ($\tau(1\mu m) = 2.1 \times \tau(2.5\mu m)$ at $L_S$ 165° whereas $\tau(1\mu m) = 2.6 \times \tau(2.5\mu m)$ between $L_S$ 96° and $L_S$ 118°). Such an increase of optical depth is consistent with the numerous local and regional dust storms observed in the northern latitudes after $L_S$ = 130° [Cantor et al., 2001]. Neumann et al. [2003] also observe an increase of the frequency of clouds at high northern latitudes after midsummer with MOLA. We have compared reflectance spectra of ice-covered regions observed with the same incidence angle (~ 72°) at $L_S$ 115° and $L_S$ 165°. Applying the aerosol correction method to these observed spectra with the optical thickness inferred above the dark regions results in a very similar reflectance spectrum of ice covered regions (Figure 17, top), which are dominated by large grained water ice. Several spectral regions (Figure 17, inserts) depend on the temperature of the ice [Grundy and Schmitt, 1998]. The observed spectral evolution in these regions is consistent with the decrease in temperature from 225 K at $L_S$ 115° to 180 K at $L_S$ 165° measured by TES [Kieffer and Titus, 2001]. The similarity of spectra at $L_S$ 115° and $L_S$ 165° indicates that no water frost deposition occurred over permanent ice at 77° N, 90° E up to one month before the fall equinox. The late summer darkening of ice patches observed by Hale et al. [2005] can be attributed to an increase in the optical thickness of aerosols, which could vary from one Mars year to the next.

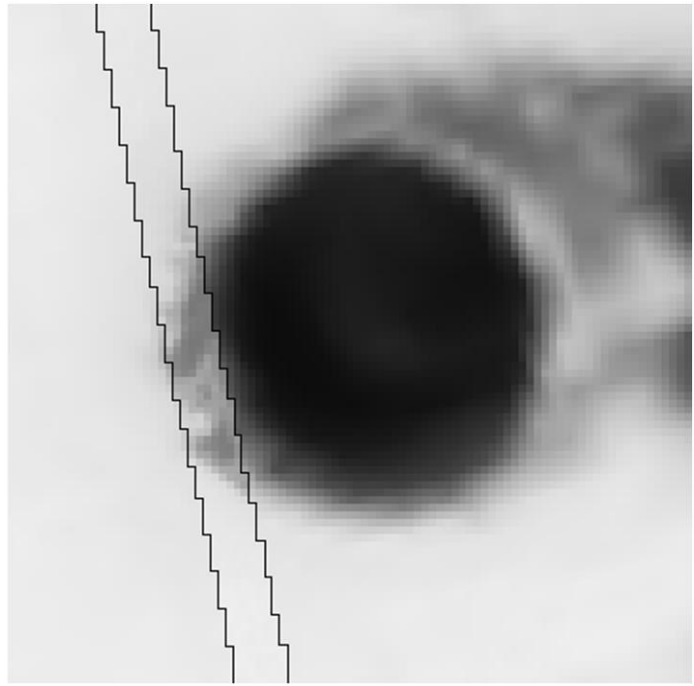

*Figure 16. High-resolution track at $L_S$ = 165° (IFOV ~ 400 m) superimposed to a map at $L_S$ = 118° of the crater bowl. The map represents the band depth at 1.5 μm, similar to Figure 6, from white (no ice) to black (band depth of 70% at 1.5 μm).*

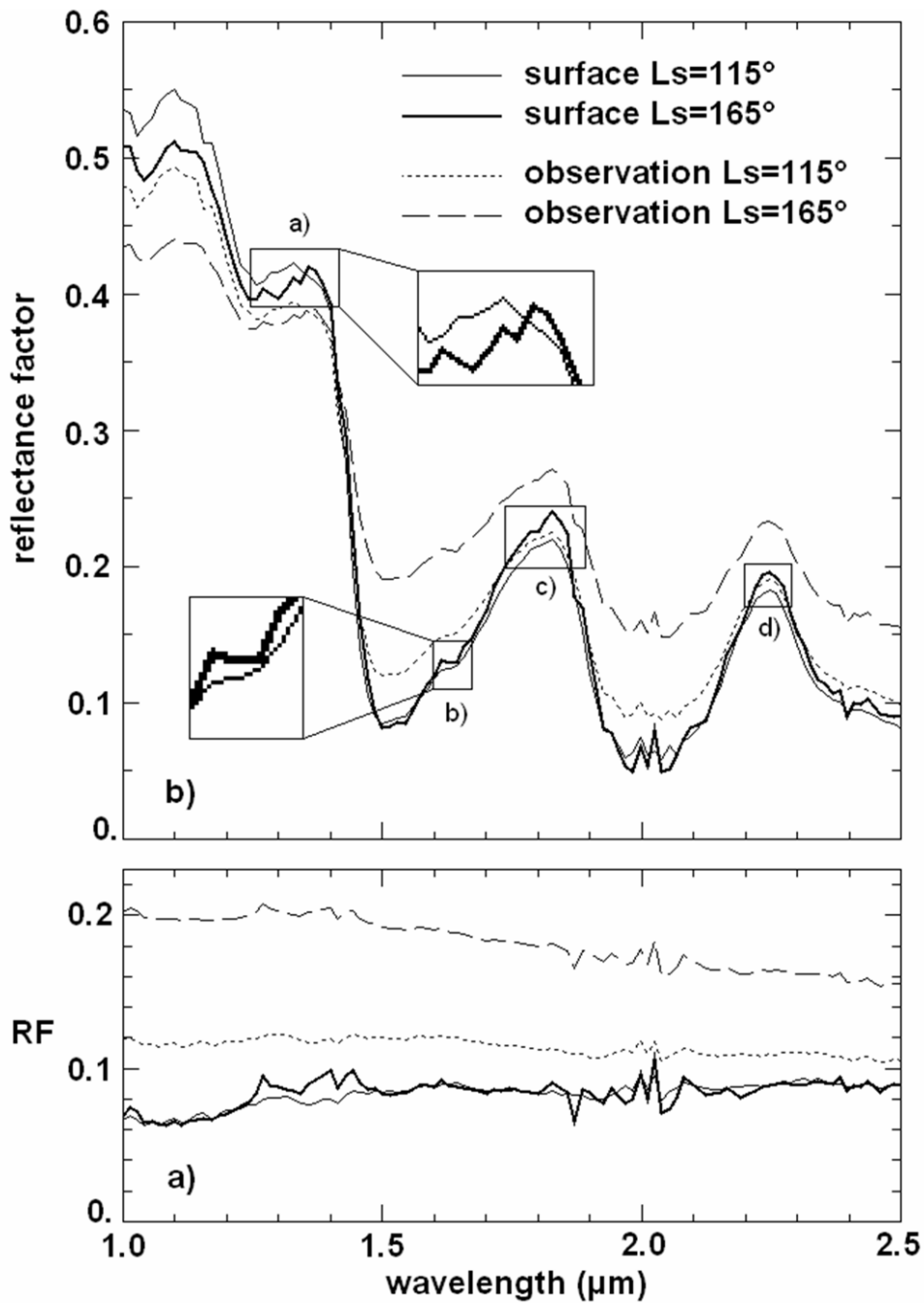

*Figure 17. Comparison between $L_S = 115°$ (#11) and $L_S = 165°$ (#15) for $i = 72°$. Dotted lines: observations at $L_S = 115°$. Dashed lines: observations at $L_S = 165°$. The reflectance spectra of the surface after removing aerosol contributions are displayed as a thin solid line ($L_S = 115°$) and a thick solid line ($L_S = 165°$). A large optical depth ($\tau$ (1 µm) ~ 0.85) can be derived from observations of a dark ice-free region at $L_S = 165°$ (Figure 17a). The reflectance spectra of ice-covered surfaces are very similar at $L_S = 115°$ and $L_S = 165°$ once the large aerosol contribution is removed, which indicates that no frost deposition has happened before $L_S = 165°$ (Figure 17b). The evolution of $H_2O$ ice spectra with temperature has been studied in the laboratory [Grundy and Schmitt, 1998]. The spectral evolution of the ice-covered surface within the H2O ice spectral bands, in particular near 1.35 µm and near 1.65 µm (insets), is consistent with the decrease of the surface temperature from 225 K at $L_S$ 115° to 180 K at $L_S$ 165° measured by TES [Kieffer and Titus, 2001].*

### 3.3 Influence of selected properties of dust aerosols

We have tested our method with the single scattering albedo obtained by Ockert-Bell et al. (~ 0.955 between 1µm and 2.5µm). We retrieve similar surface spectra and relative variation of the optical depth with wavelength. All optical depths increase by ~ 7%. This increase is expected as the single scattering albedo represents the ratio between scattered photons and interacting (scattered or absorbed) photons.

We have selected the phase function of aerosols derived by Ockert-Bell et al. [1997]. These authors obtain an asymmetry parameter g of 0.63 between 1 µm and 2.5 µm for a single-lobed Henyey-Greenstein – HG – function, relying on the size distribution of aerosols obtained by Pollack et al. [1995]. Other size distributions of aerosols such as that observed by Clancy et al. [2003] can modify the phase function and its wavelength dependence. Approaches which do not rely on a HG function have been implemented so as to model phase functions for wavelength shorter than 1µm [Tomasko et al., 1999; Markiewicz et al., 1999]. Tomasko et al. [1999, Figure 21] derived a phase function for a wavelength of 0.965µm with an asymmetry parameter of 0.7 and a higher diffraction peak than would result from a HG distribution (it presents similarities with a HG function with g=0.83). We have performed Monte-Carlo simulations using this phase function (and the corresponding single scattering albedo, 0.937) so as to reproduce observations of a dark ice-free region (Figures 11 and 13). The derived surface spectrum and relative variation of the optical depth with wavelength are very similar to those obtained with a HG distribution with g = 0.63. However, all resulting optical depths are multiplied by 1.5 (±2%). This is expected as the phase function of Tomasko et al. [1999] has a higher diffraction peak. The photons undergoing such interactions continue in nearly the same direction as those which do not interact. Therefore we need to increase the number of interactions (i.e. the optical depth) to recover a similar fraction of photons which are scattered in markedly different directions. Consequently, if the real phase function significantly varies in our spectral range, the retrieved wavelength dependence of the optical depth may be modified.

This analysis demonstrates that the recovery of surface reflectance spectra is robust with respect to the choice of assumptions on aerosol scattering properties.

### 4 Conclusions:

We have developed a model of aerosols based on Monte-Carlo methods which makes it possible to assess both aerosol extinction (which lowers the apparent albedo) and aerosol backscattering (which increases the apparent albedo) by simulating the path of a large number of incoming solar photons. This type of models can therefore be used for large optical depths and at high solar incidence angles, when multiple scattering plays a major role. The scattering properties of aerosols in the near infrared (1µm – 2.5µm) are modeled by a Henyey-Greenstein phase function with an asymmetry parameter of 0.63 independent of λ. A single scattering albedo which range from 0.971 to 0.976 has been inferred from OMEGA observations of optically thick aerosols. As a first approximation, we selected a Lambert scattering hypothesis for the surface. When observations at different geometries are available, it is then possible to determine both the aerosol thickness and the surface reflectance factor at each wavelength from the model.

We have tested this approach at high latitudes over both ice-free and ice covered regions. The same regions have been observed at close time intervals by OMEGA over a wide range of high incidence angles during summer. Furthermore, the reflectance factor for the large-grained permanent water ice ranges from a few % within major absorption bands (1.5 µm, 2 µm) to more than 50% in the continuum, while that of dust covered regions ranges from ~ 10% to ~ 30% from 1 µm to 2.7 µm. It was possible to recover consistent surface reflectance spectra for both ice-covered and ice free regions with a wavelength dependence of the optical thickness of aerosols which is within the range of previous estimates [Drossart et al., 1991; Clancy et al., 2003]. With the nadir pointing geometry which is the baseline for OMEGA observations, the variations of the apparent reflectance with the incidence angle (and phase angle) are mostly related to the contributions of aerosols, non lambertian surface scattering properties playing only a minor role.

For ice-covered regions, aerosol contributions increase the reflectance factor within absorption bands and decrease the reflectance factor in the continuum; hence the spectral contrast is reduced. The surface reflectance spectra of ice-covered regions after removing aerosol contributions are satisfactorily modeled by combining different sizes of ice grains with some dust contamination, which further validates our approach.

The spectra of dark regions which could be recovered at $L_S$ 110° and 118° from observations at different incidence angles are very similar. Assuming that the reflectance spectrum of these regions remains the same throughout summer, it is possible to determine the evolution of the aerosol optical thickness even when only one incidence angle is available. At 90° E and 77° N, it varies by a factor 3 during summer between $\tau(1\mu m)=0.85$ and $\tau(1\mu m)=0.28$ (±0.05). It is interesting to note that no frost deposition is observed in this area up to one month before the fall equinox. The brightening of the ice patches observed during early summer by Bass et al. [2000] can be predominantly attributed to a decrease in surface dust contamination associated with the sublimation of $H_2O$ frost.

Our Monte-Carlo model has been validated on a difficult test case (strong spectral contrasts, large incidence angles). It can now be applied to remove aerosol contributions from all spectra observed by OMEGA for which the aerosol optical thickness in the near IR is known. When overlapping observations by TES are available, it can be inferred from the optical thickness at 9 µm. A direct determination is possible from the OMEGA data itself if several observation geometries are available within a short time interval. This is in particular the case for the OMEGA observations implementing spot pointing. This approach is therefore of interest for CRISM/MRO, which will implement spot pointing as a nominal observation strategy [McGuire et al., 2006; Wolff et al., 2006].